\documentclass{article}
\usepackage{latexsym}
\usepackage{amsmath}
\usepackage{amsfonts}
\usepackage{amssymb}

\newcommand{\ms}{\medskip}

\newcommand{\noi}{\noindent}
\newcommand{\ra}{\rightarrow}
\newcommand{\bea}{\begin{eqnarray}}
\newcommand{\eea}{\end{eqnarray}}
\newcommand{\ol}{\overline}
\newcommand{\gr}{Groenewold}
\newcommand{\vh}{Van~Hove}
\newcommand{\vn}{Von~Neumann}
\newcommand{\q}{{\cal Q}}

\newcommand{\h}{{\cal H}}
\newcommand{\f}{{\cal F}}

\newcommand{\oo}{{\cal O}}
\newcommand{\s}{{\cal S}}

\newcommand{\pa}{Poisson algebra}
\newcommand{\pb}{Poisson bracket}
\newcommand{\la}{Lie algebra}
\newcommand{\lsa}{Lie subalgebra}
\newcommand{\sa}{subalgebra}

\newcommand{\T}{{T^*\!\,}}

\def\endproof{\hfill $\Box$}

\def\r{{\bf R}}

\newtheorem{thm}{Theorem}
\newtheorem{lem}{Lemma}
\newtheorem{cor}[thm]{Corollary}
\newtheorem{prop}[thm]{Proposition}
\newtheorem{defn}{Definition}

\def\f #1,#2.{\textstyle{#1\over #2}}

\def\sp{{\rm span}}

\begin{document}


\title{On the \gr -\vh\ problem for $\r^{2n}$\footnotetext{E-mail:
gotay@math.hawaii.edu \ \ Home Page: www.math.hawaii.edu/\~{}gotay}}

\author{{\bf Mark J. Gotay} \\ 
\\  Department of Mathematics \\ University of Hawai`i \\ 2565 The
Mall \\ Honolulu, HI 96822  USA} 

\date{September 15, 1998 \\ (Revised October 14, 1998)}

\maketitle


\begin{abstract}
We discuss the \gr -\vh\ problem for
$\r^{2n}$, and completely solve it when $n=1$. We rigorously show that
there exists an obstruction to quantizing the \pa\ of polynomials on
$\r^{2n}$, thereby filling a gap in \gr 's original proof.  
Moreover, when $n=1$ we
determine the largest Lie subalgebras of polynomials which can be
consistently quantized, and explicitly construct all their possible
quantizations.
\end{abstract}
 

\begin{section}{Introduction}

In 1946 Groenewold \cite{Gr} presented a remarkable result which
states that one cannot consistently quantize the Poisson algebra of
all polynomials in the positions
$q^i$ and momenta
$p_i$ on ${\bf R}^{2n}$ as symmetric operators on some Hilbert space
$\h,$ subject to the requirement that the
$q^i$ and  $p_i$ be irreducibly represented. Van~Hove subsequently
refined \gr's result \cite{vH1}. Thus it is {\it in principle}
impossible to quantize---by {\it any\/} means---every classical
observable on ${\bf R}^{2n}$, or even every polynomial observable, in
a way consistent with Schr\"odinger quantization (which, according to
the Stone-\vn\ theorem, is the import of the irreducibility
requirement on the $q^i$ and $p_i$). At most one can consistently
quantize certain Lie subalgebras of observables, for instance
polynomials which are at most quadratic, or observables which are
affine functions of the  momenta.

This is not quite the end of the story, however; there are two loose
ends which need to be tied up. The first is that there is a technical
gap in \gr 's proof \cite{GGT}. This gap has been filled in
\cite{vH1} (see also \cite{a-m}) by means of a certain functional
analytic assumption. Although ``small,'' this gap is nevertheless
vexing, and its elimination in this manner is not entirely
satisfactory. Second, in the absence of such a polynomial
quantization, it is important to determine the largest Lie
subalgebras of polynomials that can be consistently quantized along
with their quantizations. While some results are known along these
lines, this program has not yet been fully carried out.

In this paper we consider the \gr -\vh\ problem for
$\r^{2n}$. We present two variants of \gr's theorem (``strong''
and ``weak''); the weak one is the version that \gr\ actually proved,
while the strong one is the result referred to above.
We then show that the strong version follows from the
weak one \emph{without} introducing extra hypotheses.
Thus we fill the gap in \gr 's proof. Moreover, when $n=1$ we
determine the largest quantizable Lie subalgebras of polynomials and
explicitly construct all their possible quantizations.

To make the presentation self-contained, we include a detailed
discussion of previous work on the \gr -\vh\ problem.

We wish to thank H. Grundling for valuable discussions,
including his help in pointing out and correcting an error in a
previous version of this manuscript. 
This research was supported in part by NSF grant DMS 96-23083.

\end{section}


\begin{section}{Background}

Let $P(2n)$ denote the Poisson algebra of polynomials on $\r^{2n}$
with Poisson bracket
\[ \{f,g\} = \sum_{k=1}^n\left[\frac{\partial f}{\partial
p_k}\frac{\partial g}{\partial q^k} -
\frac{\partial g}{\partial p_k} \frac{\partial f}{\partial
q^k}\right].\]

\noi $P(2n)$ contains several distinguished \lsa s: the
\emph{Heisenberg algebra}
\[{\rm h}(2n) = \sp \big\{1,q^i,p_i\,|\,i=1,\ldots,n\big\};\] 

\noi the \emph{symplectic algebra}
\[{\rm sp}(2n,\r) = \sp \big\{q^i q^j,q^i
p_j,p_i p_j\,|\,i,j=1,\ldots,n\big\};\] 

\noi the \emph{extended} (or \emph{inhomogeneous}) \emph{symplectic
algebra}
\[{\rm hsp}(2n,\r) = \sp \big\{1,q^i,p_j,q^i q^j,q^i 
p_j,p_i p_j\,|\,i,j=1,\ldots,n\big\},\] 

\noi which is the semidirect product of h(2$n$) with sp$(2n,\r)$; and
the \emph{coordinate} (or \emph{posi\-tion}) \emph{algebra}
\[C(2n) = \left\{\sum_{i=1}^n f^i\!(q)p_i + g(q)\right\},\]

\noi where $f^i$ and $g$ are polynomials. All of these will play an
important role in our development.

Let
$P^k(2n)$ denote the subspace of polynomials of 
degree at most $k$, and
$P_k(2n)$ the subspace of homogeneous polynomials of degree
$k$. Then $P^1(2n) = {\rm h}(2n)$, $P_2(2n) = {\rm sp}(2n,\r)$, and
$P^2(2n) = {\rm hsp}(2n,\r).$ When
$n$ is fixed, we simply write $P = P(2n)$, etc.
\ms 

We now state what it
means to ``quantize'' a Lie algebra of polynomials on
$\r^{2n}$. Throughout, the Heisenberg algebra h$(2n)$ is regarded as a
``basic algebra of observables,'' cf. \cite{go98}.

\begin{defn}$\,\,$ Let $\oo$ be a Lie
subalgebra of $P(2n)$ containing the Heisenberg algebra {\rm
h}$(2n)$. A {\rm quantization\/} of\/
$\oo$ is a linear map
$\q$ from $\oo$ to the linear space {\rm Op({\it D})} of
symmetric operators which preserve a fixed dense domain $D$ in some
separable Hilbert space
$\h$, such that for all $f,g \in \oo$,

\begin{description}
\item \rule{0mm}{0mm}
\begin{enumerate}
\vspace{-5ex}
\item[{\rm (Q1)}]  ${\cal Q}(\{f,g\}) =
\frac{i}{\hbar}[{\q}(f),{\q}(g)]$,
\vskip 6pt
\item[{\rm (Q2)}]  ${\cal Q}(1) = I$, 
\vskip 6pt
\item[{\rm (Q3)}] if the Hamiltonian vector field $X_f$ of
$f$ is complete, then $\q(f)$ is essentially self-adjoint on $D$,
\vskip 6pt
\item[{\rm (Q4)}] $\q$ represents ${\rm h}(2n)$ irreducibly, and 
\vskip 6pt
\item[{\rm (Q5)}] $D$ contains a dense set of separately analytic
vectors for the standard basis of $\q({\rm h}(2n)).$\footnote{A
vector $\psi$ is \emph{analytic} for an operator $A$ provided
$A^n\psi$ is defined for all integers $n \geq 0$, and 
\[\sum_{n=0}^\infty \frac{1}{n!}\| A^n\psi\|t^n < \infty\]

\noi for some $t > 0.$ $\psi$ is \emph{separately analytic} for a
collection of operators if it is analytic for each operator in the
collection.}
\end{enumerate}
\end{description}
\label{def:q}
\end{defn}

\vskip -1ex
We briefly comment on these conditions; a full exposition
along with detailed motivation is given in \cite{go98}.  Condition
(Q1) is Dirac's famous ``\pb\ $\ra$ commutator'' rule; here $\hbar$ is
Planck's reduced constant. The second condition reflects the fact
that if an observable $f$ is a constant $c$, then the probability of
measuring $f=c$ is one regardless of which quantum state the system is
in. Regarding (Q3), we remark that in contradistinction with \vh\
\cite{vH1}, we do not confine our considerations to only those
classical observables whose Hamiltonian vector fields are complete.
Rather than taking the point of view that ``incomplete'' classical
observables cannot be quantized, we simply do not demand that the
corresponding quantum operators be essentially self-adjoint
(``e.s.a.''). 

(Q4) and (Q5) emphasize the fundamental role of the
Heisenberg algebra. The technical condition (Q5) guarantees the
integrability of the Lie algebra representation $\q({\rm h}(2n))$ on
$D$ \cite{f-s}.  (There do exist nonintegrable representations of the
Heisenberg algebra, e.g. \cite[p. 275]{ReSi}; however, none of them
seem to have physical significance. (Q5) thus serves to eliminate
these ``spurious'' representations.) By virtue of the Stone-\vn\
theorem, (Q5) along with the irreducibility criterion (Q4) imply that
$\q({\rm h}(2n))$ is unitarily equivalent to a restriction of the {\it
Schr{\"o}dinger quantization\/} $d\Pi$:
\begin{equation} q^i \mapsto q^i,\;\;\; p_j \mapsto -i\hbar \,
\partial /{\partial q^j},\;\;\; \mbox{and}\;\;\; 1 \mapsto I
\label{eq:srep}
\end{equation} 

\noi on the Schwartz space $\s (\r^n) \subset L^2({\r^{n}})$.
Indeed, by (Q5)
$\q({\rm h}(2n))$ can be integrated to a representation
$\tau$ of the Heisenberg group ${\rm H}(2n)$ which, according to
(Q4), is irreducible. The Stone-\vn\
theorem then states that this representation of H$(2n)$ is unitarily
equivalent to the Schr\"odinger representation $\Pi$, and hence $\tau
= U\Pi U^{-1}$ for some unitary map $U:L^2(\r^n) \ra \h$.
Consequently, $\q(f) = U\overline{d\Pi(f)} U^{-1} \!\restriction \! D$
for all $f \in {\rm h}(2n)$ \cite[Cor. 1]{f-s}, where the bar
denotes closure. 
It now follows from (\ref{eq:srep}), the invariance of the domain $D$,
and Sobolev's lemma that $U^{-1}D \subseteq {\mathcal S}(\r^n)$, so
that $U^{-1}\q U$ is the restriction of $d\Pi$ to $U^{-1}D$.

Finally, in \cite{go98} there is a sixth criterion that a
quantization must satisfy in general, viz. that $\q$ be faithful when
restricted to the given basic algebra of observables. In the case of
the Heisenberg algebra, however, this holds automatically in view
of (Q1) and (Q2).

\end{section}


\begin{section}{The Weak No-Go Theorem}

In the next two sections we argue that there are no 
quantizations of $P(2n)$. Extensive discussions can be found in
\cite{a-m,Ch1,Fo,GGra,GGT,go80,Gr,GS,vH1}. We shall state the
main results for $\r^{2n}$ but, for convenience, usually prove them
only for $n=1$. The proofs for higher dimensions are immediate
generalizations of these. 

We begin by observing that there {\em does\/} exist a quantization
$d\varpi$ of hsp($2n,\r$). It is given by the familiar formul{\ae}
\begin{eqnarray*} d\varpi(q^i)  = q^i, & d\varpi(1) = I, &
d\varpi(p_j) = -i\hbar
\frac{\partial}{\partial q^j}, 
\label{eq:p1} 
\end{eqnarray*}
\vskip -1ex
\begin{eqnarray}  d\varpi(q^iq^j) = q^i q^j, &  d\varpi(p_ip_j) =
-\hbar^2 {\displaystyle \frac{\partial ^2}{\partial q^i \partial
q^j}},
\label{eq:p2}
\end{eqnarray}
\vskip -1ex
\begin{eqnarray}
& d\varpi(q^ip_j) =
-i\hbar {\displaystyle
\left (q^i \frac{\partial}{\partial q^j} + \frac{1}{2}\delta^i_j\right
)}, &
\label{eq:p3}
\end{eqnarray}

\vskip 1ex

\noi on the domain $\s (\r^n) \subset L^2(\r^n).$ Properties
(Q1)--(Q3) are readily verified. Property (Q4) follows automatically
since the restriction of
$d\varpi$ to
$P^1$ is just the Schr\"od\-inger representation. For (Q5) we recall
that the Hermite functions $h_{k_1 \cdots k_n}(q^1,\ldots,q^n) =
h_{k_1}(q^1) \cdots h_{k_n}(q^n)$, where
\begin{equation*}
\label{eq:hermite}
h_k(q) = e^{q^2/2} \frac{d^k}{dq^k} e^{-q^2}
\end{equation*}

\noi for $k = 0,1,2,\ldots$, form a dense set of
separately analytic vectors for $d\varpi(P^1)$. As these functions
are also separately analytic vectors for $d\varpi(P_2)$
\cite[Prop.~4.49]{Fo}, the operator algebra
$d\varpi(P^2)$ is integrable to a unique representation $\varpi$ of
the universal cover ${\widetilde{\rm{HSp}}}(2n,\r)$ of the extended
(or inhomogeneous) symplectic group HSp$(2n,\r)$ (thereby justifying
our notation ``$d\varpi$'').\footnote{\,This representation actually
drops to the double cover of HSp$(2n,\r)$, but we do not need this
fact here.}
$\varpi$ is known as the ``extended metaplectic representation''; 
detailed discussions of it may be found in \cite{Fo,GS}.

We call $d\varpi$ the ``extended metaplectic quantization.'' It has
the following crucial property.
\begin{prop} The extended metaplectic quanti\-zation is the {\em
unique\/}  quantization of 
${\rm hsp}(2n,\r)$ which exponentiates to a
unitary representation of \linebreak[4] $\widetilde{\rm HSp}(2n,\r)$.
\label{prop:unique}
\end{prop}

By ``unique,'' we mean up to unitary equivalence and restriction of
representations.

\ms

\noi {\it Proof.} Suppose $\q$ were another such quantization of
${\rm hsp}(2n,\r)$ on some domain $D$ in
a Hilbert space
$\h$. Then
$\q\big({\rm hsp}(2n,\r)\big)$ can be integrated to a representation
$\tau$ of ${\widetilde{\rm HSp}}(2n,\r)$, and (Q4) implies that
$\tau$, when restricted to H$(2n)
\subset {\widetilde{\rm HSp}}(2n,\r)$, is irreducible. The Stone-\vn\
theorem then states that this representation of H$(2n)$ is unitarily
equivalent to the Schr\"odinger representation, and hence $\tau =
U\varpi U^{-1}$ for some unitary map $U:L^2(\r^n) \ra \h$ by
\cite[Prop.~4.58]{Fo}. Consequently, $\q(f) = U\overline{d\varpi(f)}
U^{-1} \! \restriction \! D$ for all $f \in {\rm hsp}(2n,\r)$. 
Arguing as in the discussion following Definition~\ref{def:q}, we see
that $U^{-1}\q U$ is in fact the \emph{restriction} of $d\varpi$ to
$U^{-1}D \subseteq {\mathcal S}(\r^n).$
\endproof

\ms 

The existence of an obstruction to quantization now follows from
\begin{thm}[Weak No-Go Theorem] The extended metaplectic quantization
of\/
$P^2$ cannot be extended beyond $P^2$ in $P$.
\label{thm:weak}
\end{thm}

Since $P^2$ is a maximal Lie subalgebra of $P$ \cite[\S 16]{GS}, (Q1)
implies that any quantization which extends $d\varpi$ must be
defined on all of $P$. Thus we may restate this as: {\em There exists
no quantization of\/
$P$ which reduces to the extended metaplectic quantization on
$P^2.$}
\ms

\noi {\it Proof.} Let $\q$ be a quantization of $P$ which
extends the metaplectic quantization of $P^2$. As noted previously,
we may assume that the domain $D \subseteq {\mathcal S}(\r^n)$. We
will show that a contradiction arises when cubic polynomials are
considered.

Take $n=1$. By inspection of (\ref{eq:srep})--(\ref{eq:p3}) we see
that the ``\emph{\vn\ rules}'' 
\begin{equation}
\q(q^2)=\q(q)^2,\;\; \q(p^2)=\q(p)^2, 
\label{eq:vnrr2n1}
\end{equation}
\begin{equation}
\q(q\,\!p) = {\textstyle \frac{1}{2}}\big(\q(q) \q(p) + \q(p)
\q(q)\big), 
\label{eq:vnrr2n2}
\end{equation}

\vskip 1ex

\noi hold. These in turn lead to higher degree \vn\ rules
\cite{Ch1,Fo}.

\begin{lem} For all real-valued polynomials $r$,
\[\q\big(r(q)\big) = r\big(\q(q)\big),\;\;\;\q\big(r(p)\big)=
r\big(\q(p)\big),\]
\[\q\big(r(q)p\big) = {\textstyle
\frac{1}{2}}\big[r\big(\q(q)\big)\q(p)+\q(p)r\big(\q(q)\big)\big],\]
\[\q\big(qr(p)\big) = {\textstyle
\frac{1}{2}}\big[\q(q)r\big(\q(p)\big)+r\big(\q(p)\big)\q(q)\big].\]
\label{lem:morevnrs}
\end{lem}

\vskip -3ex

\noi {\it Proof.} We illustrate this for $r(q) = q^3$. The other rules
follow similarly using induction. Now $\{q^3,q\} = 0$ whence by (Q1)
we have
$[\q(q^3),\q(q)] = 0$. Since also
$[\q(q)^3,\q(q)] = 0$, we may write $\q(q^3) =
\q(q)^3 + T$ for some operator $T$ which (weakly) commutes with
$\q(q)$. We likewise have using (\ref{eq:vnrr2n1})
\[ [\q(q^3),\q(p)] =  -i\hbar\,
\q(\{q^3,p\}) 
 =  3i\hbar\,\q(q^2) =  3i\hbar\,\q(q)^2 =
[\q(q)^3,\q(p)]\]

\noi from which we see that $T$ commutes with $\q(p)$ as well.
Consequently, $T$ also commutes with $\q(q)\q(p)+ \q(p)\q(q)$. But
then from (\ref{eq:vnrr2n2}),
\begin{eqnarray*}\q(q^3) & = & {\textstyle \frac{1}{3}}\,
\q\big(\{pq,q^3\}\big) = {\textstyle
\frac{i}{3\hbar}}\,[\q(pq),\q(q^3)] \\ & =& {\textstyle
\frac{i}{3\hbar}}\,\left[{\textstyle
\frac{1}{2}}\big(\q(q)\q(p) +
\q(p)\q(q)\big),\q(q)^3 + T\right] \\ & = & {\textstyle
\frac{i}{6\hbar}}\,[\q(q)\q(p) +
\q(p)\q(q),\q(q)^3] 
 = \q(q)^3.  
\end{eqnarray*}
 \vskip -5ex \hfill $\bigtriangledown$

\ms

\ms

With this lemma in hand, it is now a simple matter to prove the no-go
theorem. Consider the classical equality 
\[{\textstyle \frac{1}{9}}\{q^3,p^3\}= {\textstyle
\frac{1}{3}}\{q^2p,p^2q\}.\]

\noi Quantizing and then simplifying this, the formul{\ae} in
Lemma~\ref{lem:morevnrs} give
\[\q(q)^2\q(p)^2 - 2i\hbar\q(q)\q(p) - {\textstyle
\frac{2}{3}}\hbar^2 I\]

\noi for the L.H.S., and 
\[\q(q)^2\q(p)^2 - 2i\hbar\q(q)\q(p) - {\textstyle \frac{1}{3}}\hbar^2
I\]

\noi for the R.H.S., which is a contradiction.
\endproof

\end{section}


\begin{section}{The Strong No-Go Theorem}

In \gr's paper \cite{Gr} a stronger result was claimed: his assertion
was that there is no quantization of
$P$, period. This is \emph{not} what Theorem~\ref{thm:weak} states.
For if $\q$ is a quantization of $P$, then while of course
$\q(P^1)$ must coincide with Schr\"odinger
quantization, it is not obvious that $\q$ need be the extended
metaplectic quantization when restricted to
$P^2$. Referring to Proposition~\ref{prop:unique}, the problem is
that $\q(P^2)$ is not
\emph{a priori} integrable; (Q5) only guarantees that
$\q(P^1)$ can be integrated. {}From a different 
point of view, the problem lies in \emph{deducing} the relations
(\ref{eq:p2}) and (\ref{eq:p3}) (or, what amounts to the same
thing, the \vn\ rules (\ref{eq:vnrr2n1}) and (\ref{eq:vnrr2n2})) from
the quantization axioms (Q1)--(Q5) and the properties of the extended
symplectic algebra alone. 

\vh\ supplied an additional assumption guaranteeing the
integrability of $\q(P^2)$, which in particular implies: If the
Hamiltonian vector fields of $f,\,g$ are complete and $\{f,g\} = 0$,
then $\q(f)$ and $\q(g)$ {\em strongly\/} commute
\cite{vH1}.\footnote{\,Recall that two e.s.a.\  operators {\it
strongly commute\/} iff their spectral resolutions commute, cf.
\cite[\S VIII.5]{ReSi}. Two operators $A,\,B$ {\it weakly commute\/}
on a domain $D$ if they commute in the ordinary sense, i.e., $[A,B]$
is defined on $D$ and vanishes.} This
assumption is used to derive the relations
(\ref{eq:p2}) and (\ref{eq:p3}) in \cite{a-m, Ch1}. It is also possible to enforce
the integrability of $\q(P^2)$ in a more direct manner \cite{GGT}. 

We now show that \vh 's assumption is unnecessary; in fact,
we may establish the integrability of $\q(P^2)$ directly, 
via the following generalization of Proposition~\ref{prop:unique}.
\begin{prop}
\label{prop:fix}
Let $\q$ be a quantization of $P^2$ on a dense invariant domain $D$
in a Hilbert space $\mathcal H$. Then there is a unitary
transformation $U: L^2(\r^n) \ra \mathcal H$  
such that
$\q(f) = Ud\varpi(f) U^{-1}\!\restriction\! D$ for
all $f\in P^2.$
\end{prop}

Thus, up to unitary equivalence, $\q$ 
must be either $d\varpi$ or a
restriction thereof. As such, $\q(P^2)$ must be integrable and,
consequently, \vh 's strong commutativity assumption holds for
elements of $P^2$. 

Before giving the proof, we establish two technical lemmas. 
\begin{lem} Let $A$
be an e.s.a. operator on a Hilbert space, and $B$ a
closable operator, both of which have a common dense invariant domain
$D$. Suppose that $D$ consists of analytic vectors for $A$,
and that $A$ {\rm (}weakly{\rm )} commutes with $B$. Then $\exp(i
\bar A)$ {\rm (}weakly{\rm )} commutes with $\bar B$ on $D$.
\label{lem:fa1} 
\end{lem}

\noi {\it Proof.} Recall that as $\psi \in D$ is analytic for $A$,
\[e^{i \bar A} \psi = \sum_{k=0}^\infty \frac{1}{k!}(iA)^k \psi =:
\phi.\]

\noi Define the partial sums
\[\phi_K = \sum_{k=0}^K \frac{1}{k!}(iA)^k \psi \in D;\]

\noi then using the (weak) commutativity of $A$ and $B$,
\[B\phi_K = \sum_{k=0}^K \frac{1}{k!}(iA)^k B\psi \in D. \]

\noi Since $B\psi \in D$ is analytic for $A$, the sequence $B\phi_K$
converges:
\[\chi := \lim_{K \ra \infty} B\phi_K = e^{i \bar
A} B\psi = e^{i \bar A} {\bar B}\psi.\]

\noi But $\bar B$ is closed, hence $\phi = \lim_{K \ra \infty}
\phi_K$ is in the domain of $\bar B$ and $\chi = {\bar B}\phi$, i.e.
\[e^{i \bar A}{\bar B} \psi = {\bar B} e^{i \bar A}\psi\]

\noi for all $\psi \in D$. \hfill $\bigtriangledown$
\begin{lem} Let $B$ be a closable operator. If a bounded operator $T$
{\rm (}weakly{\rm )} commutes with
$\bar B$ on
$D(B)$, then they also commute on $D(\bar B)$.
\label{lem:fa2} 
\end{lem}

\noi {\it Proof.} If $\psi \in D(\bar B)$, then from the definition of
closure there exists a sequence $\{\psi_k\}$ in $D(B)$
with $\psi_k \ra \psi$ such that $B\psi_k \ra 
{\bar B}\psi.$ Because the operator $T$ is
continuous, 
\begin{equation*}
T{\bar B} \psi  =  T\lim_{k \ra \infty}B\psi_k
= \lim_{k \ra \infty}TB\psi_k = \lim_{k \ra
\infty}{\bar B}T\psi_k
\end{equation*}

\noi as $T$ commutes with $\bar B$ on $D(B)$. Again applying
the definition of closure to the sequence $\{T\psi_k\}$ in
$D(\bar B)$, we get that $\lim_{k \ra \infty}T\psi_k =
T\psi \in D(\bar B)$ and

\[{\bar B}T\psi = \lim_{k \ra
\infty}{\bar B}T\psi_k = T{\bar B} \psi\] 

\noi for every $\psi \in D(\bar B)$. \hfill $\bigtriangledown$

\ms

\noi {\it Proof of Proposition \ref{prop:fix}.} Let $\q$ be a
quantization of $P^2$. As discussed earlier, we
may assume that $\q(P^1)$ is the Schr\" odinger representation
(\ref{eq:srep}) on $L^2(\r^n)$, and that the domain $D \subseteq
{\mathcal S}(\r^n)$. Again taking
$n =1$, we will prove by brute force that the \vn\ rules
(\ref{eq:vnrr2n1}) and (\ref{eq:vnrr2n2}) hold. 

We begin by determining $\q(q^2)$. 
Set $\Delta = \q(q^2) - \q(q)^2$. We readily verify that
$[\Delta,\q(q)] = 0$ and  $[\Delta,\q(p)] = 0$ on $D$.
Now let
$D_\omega \subseteq D$ be the space of separately analytic vectors for
$\q(q)$ and $\q(p)$; by (Q5) we have that $D_\omega$ is dense.
According to \cite[Prop. 1]{f-s},  $\q(P^2)$ leaves
$D_\omega$ invariant and consequently so does $\Delta$. By 
\cite[\S X.6, Cor.~2]{ReSiII}
$\q(q)\!\restriction\! D_\omega$ is e.s.a; moreover, $\Delta_\omega :=
\Delta \!\restriction\! D_\omega$ is symmetric and hence closable.
Upon taking $A = \q(q)\!\restriction\! D_\omega$ and $B =
\Delta_\omega$ in Lemma~\ref{lem:fa1}, it follows that
$\exp(i\ol{\q(q)\!\restriction\! D_\omega}) = \exp(i\ol{\q(q)})$ and
$\ol{\Delta_\omega}$ commute on
$D_\omega$. Lemma~\ref{lem:fa2} then shows that $\exp(i\ol{\q(q)})$
and $\ol{\Delta_\omega}$ commute on
$D(\ol{\Delta_\omega})$. Likewise
$\exp(i\ol{\q(p)})$ and $\ol{\Delta_\omega}$ commute on $D(\ol{
\Delta_\omega})$. But now the unbounded version of Schur's lemma
\cite[(15.12)]{r} implies that $\ol{\Delta_\omega} = EI$ for some real
constant
$E.$ Since
$\ol{\Delta_\omega}$ is the smallest closed extension of
${\Delta_\omega}$ and
$\Delta_\omega \subset \Delta \subset \bar \Delta$, it follows that
$\bar \Delta =
E I$, whence $\Delta$ itself is a multiple of the identity on
$D$. Thus $\q(q^2) = \q(q)^2 + EI$ on $D$.

An identical argument yields $\q(p^2) = \q(p)^2 + FI$ on $D.$
Quantizing the relation $4pq = \{p^2,q^2\}$ and using these
formul\ae\  then gives
\[\q(pq) = {\textstyle \frac{1}{2}}\big(\q(p)\q(q) +
\q(q)\q(p)\big)\]

\noi on $D.$ But upon quantizing $2q^2 = \{pq,q^2\}$ we find
that $E=0$. Similarly $F=0$. It follows from
(\ref{eq:srep})--(\ref{eq:p3}) that
$\q = d\varpi\!\restriction \! D$.
\endproof

\ms

Thus, up to unitary equivalence and restriction of representations,
we may as well suppose that $D = {\mathcal
S}(\r^{2n})$. If we were to take this as our starting point,
then we could reverse reverse our constructions and derive
(\ref{eq:vnrr2n1}) and (\ref{eq:vnrr2n2}) in a simpler
fashion \cite[\S 5.1]{go98}. 
 
\ms

If $\q$ were a quantization of $P$, $\q(P^2)$ must
therefore be unitarily equivalent to (a restriction of) the extended
metaplectic quantization, and this contradicts
Theorem~\ref{thm:weak}. Thus we have proven our main result:
\begin{thm}[Strong No-Go Theorem] There is no quantization
of $P$.
\label{thm:strong}
\end{thm}

Van~Hove \cite{vH1} gave a slightly different analysis using only
those observables $f\in C^\infty(\r^{2n})$ with complete Hamiltonian
vector fields, and still obtained an obstruction (but now to
quantizing all of $C^\infty(\r^{2n})$). Yet other variants of \gr 's
theorem are presented in \cite{D,GGra,Jo}. Related results can be
found in \cite{a-b}.

\end{section}


\begin{section}{Quantizable Lie Subalgebras of Polynomials}

We hasten to add that there are Lie subalgebras of $P(2n)$ other
than $P^2(2n)$ which can be quantized. For example, consider the
coordinate algebra $C(2n)$. It is straightforward to verify
that for each $\eta \in \r,$ the map
$\sigma_{\eta}:C(2n) \ra {\rm Op}\big(\s (\r^n)\big)$ given by 
\begin{equation}
\sigma_{\eta}\left(\sum_{i=1}^n f^i\! (q)p_i + g(q)\right) =
-i\hbar \sum_{i=1}^n\left(f^i\! (q)\frac{\partial}{\partial q^i} +
\left[
\frac{1}{2} + i\eta \right]\frac{\partial f^i\! (q)}{\partial
q^i}\right) + g(q)
\label{eq:sigma}
\end{equation}

\noi is a quantization of $C(2n).$ $\sigma_0$
is the familiar ``position'' or ``coordinate representation.'' The
significance of the parameter $\eta$ is explained in \cite{ADT,An}
(see also \cite{GG1}). There it is shown that while the quantizations
$\sigma_\eta$ and $\sigma_{\eta'}$ are unitarily inequivalent if
$\eta \neq \eta'$, they are related by a \emph{nonlinear}
norm-preserving isomorphism.

\begin{prop}
$C$ is a maximal \lsa\ of $P.$
\label{prop:lsa}
\end{prop}

\noi {\it Proof.}
We take $n=1.$ Suppose that $V$ were a \lsa\ of $P$ strictly
containing $C$. $V$ must contain a polynomial $h$ of the form
\[h(q,p) = f(q)p^k + \mbox{terms of degree
at most $k-1$ in $p$}\]

\noi for some $k > 1$ and some polynomial $f \neq 0$ of degree $l$.
Now both $q,p \in V$, and so by bracketing $h$ with $q$ $(k-2)$-times,
we get 
\begin{equation*}
\frac{k!}{2}f(q)p^2 +  \mbox{terms of degree at most degree 1 in
$p$} \in V.
\end{equation*}

\noi Since $C \subset V$ this implies that $f(q)p^2 \in V$. By
bracketing this expression with $p$ $l$-times, we conclude that
$p^2 \in V$. Now both $q^2, q\,\! p \in V$, so $P^2 \subset V$. The
maximality of $P^2$ implies that $V = P$, whence $C$ is maximal.
\endproof

\ms

As a consequence, any
quantization which extends $\sigma_\eta$ must be defined on all of
$P$. Thus Theorem~\ref{thm:strong} yields
\begin{cor} The quantizations $\sigma_\eta$ of\/
$C$ cannot be extended beyond
$C$ in $P$.
\label{cor:c}
\end{cor}

Furthermore a variant of Proposition~\ref{prop:fix}
(see also \cite[Thm. 8]{GG1})
yields
``uniqueness'': 
\begin{prop} 
Let $\q$ be a quantization of $\, C$ on a dense invariant domain $D$
in a Hilbert space $\mathcal H$. Then there is an $\eta \in \r$ and a
unitary transformation $U: L^2(\r^n) \ra \mathcal H$ such that
$\q(f) = U\sigma_\eta(f) U^{-1}\!\restriction\!
D$ for all $f\in C.$
\end{prop}

\noi {\it Proof.} Again set $n=1.$ As in the proof of
Proposition~\ref{prop:fix}, we may assume that $\q(P^1)$ is given
by (\ref{eq:srep}) on $L^2(\r)$ and that $D \subseteq
{\mathcal S}(\r)$. 

Just as before, we first compute that
$\q(q^2) = \q(q)^2 + EI$ on $D$ for some (real) constant $E$.

Now consider $\q(q\,\!p).$  Set
\[\Delta = \q(q\,\!p) - {\textstyle \frac{1}{2}}\big(\q(q)\q(p) +
\q(p)\q(q)\big).\]

\noi It is straightforward to verify that $\Delta$
commutes with both $\q(q)$ and $\q(p)$. The same argument based on
Lemmas~\ref{lem:fa1} and \ref{lem:fa2} and the unbounded Schur's
lemma that was used in the proof of Proposition~\ref{prop:fix} can be
applied \emph{mutatis mutandis} to give $\Delta = GI$ on $D$ for some
real constant $G$. Thus
\begin{equation}
\q(q\,\!p) = \textstyle{\frac{1}{2}}\big(\q(q) \q(p) + \q(p)
\q(q)\big) + GI
\label{eq:qp}
\end{equation}

\noi on $D$. By quantizing the
\pb\ relation $\{q\,\! p,q^2\} = 2q^2$ we find that $E=0$.
Arguing as in the proof of Lemma~\ref{lem:morevnrs}, we then compute
that on $D$
\begin{equation}
\q(q^k) = \q(q)^k.
\label{eq:qk}
\end{equation}

Next, quantizing the \pb\ relations $\{q^k p,q\} = q^k$ and
$\{q^k p,p\} = -kq^{k-1} p$ yields
\begin{equation}
[\q(q^k p),\q(q)] = -i\hbar \q(q^k) \;\;\;\mbox{ and }\;\;\;
[\q(q^k p),\q(p)] = i\hbar k \q(q^{k-1} p),
\label{eq:commutators}
\end{equation}

\noi respectively. Now consider the classical relation
$(1-k)q^k p = \{q^k p,q\,\! p\}$. Quantizing this and simplifying by
means of (\ref{eq:qp}), (\ref{eq:commutators}), and (\ref{eq:qk})
produces the recursion relation
\[\q(q^k p) = \frac{1}{1-k}\left(\q(q^k)\q(p) - k
\q(q)\q(q^{k-1} p)\right).\]

\noi Iterating this computation $(k-1)$-times gives
\[\q(q^k p) = (1-k)\q(q^k)\q(p) + k
\q(q)^{k-1}\q(q\,\! p).\]

\noi Again using (\ref{eq:qp}) and simplifying, we finally get
\[\q(q^kp)  = \q(q^k)\q(p) + k\left(G -
\frac{i\hbar}{2}\right)\q(q)^{k-1}.\]

\noi Recalling (\ref{eq:srep}) and (\ref{eq:qk}), this can be
rewritten
\[\q(q^k p) = -i\hbar \left[q^k\frac{d}{dq} +
\left(\frac{1}{2} + \frac{iG}{\hbar}\right)\frac{dq^k}{dq}\right]\]

\noi on $D$. Consolidating this with (\ref{eq:qk}), we obtain
(\ref{eq:sigma}) where  $\eta = G/\hbar.$ We thus have $\q
= \sigma_\eta \!\restriction\! D,$ as claimed.
\endproof

\ms 
 
Notice that unlike in the proof of Proposition~\ref{prop:fix}, we
cannot quantize the \pb\ relation $\{q^2,p^2\} = -4 q\,\!p$ to obtain
$G = 0$ since $p^2 \not \in C$. The fact that $G$ remains arbitrary is
mirrored by the presence of the parameter $\eta$ in
(\ref{eq:sigma}).

\ms

Thus far we have encountered two maximal \lsa s of $P$ containing
$P^1$: $P^2$ and $C$. When
$n=1$, it turns out that these are essentially the \emph{only}
such \sa s.

\begin{thm} {\rm ($n=1$)} Up to isomorphism,
$P^2$ and $C$ are the only maximal
Lie subalgebras of\/ $P$ which contain $P^1$.
\label{thm:max}
\end{thm}

\noi {\it Proof.} 
Suppose that $W$ were a
maximal \lsa\ of $P$ containing $P^1$, distinct from $P^2$. We will
show that $W$ must be isomorphic to $C$. Denote $W^k = W \cap P^k,$
etc. 

Since $W \neq P^2$ there must exist a polynomial of degree $k$,
$k > 2$, in $W$. By bracketing this polynomial $(k-2)$
times with an appropriate number of $p,q \in W$, we
obtain a nonzero polynomial $h \in W^2$. Since $P^1 \subset W$, we
may subtract off terms of degree one or less, so we may assume that
$h$ is homogeneous quadratic. By means of a rotation we may
diagonalize
$h$; thus we may further suppose that canonical coordinates have been
chosen so that $h(q,p) = ap^2 + cq^2.$ Now $\dim W_2 \neq 3$, for
otherwise
$P^2 \subset W$, and then the maximality of $P^2$ implies that $W =
P.$ We break the argument into parts, depending on whether $\dim
W_2 = 1$ or 2.

\vskip 6pt
(\emph{i}) $\dim W_2 = 1$: \ \ Then $W_2$ is
spanned by $h.$ We first claim that either $W^3 = W^2$ or $W^3 \subset
C^3$. Indeed, if
$f
\in W^3$, then the
quadratic terms of both
$\{p,f\}, \{f,q\} \in W^2$ must be proportional to $h$: $\{p,f\} = rh
+ {\rm l.d.t.}$ and
$\{f,q\} = sh + {\rm l.d.t.}$, where ``l.d.t.'' means lower
degree terms. The particular form of $h$ then implies that 
\[f(q,p) = {\textstyle \frac{1}{3}}(sap^3 + rcq^3) + {\rm l.d.t.}\]

\noi  along with $sc = 0$ and $ra=0$. Since $h \neq 0$, both $a,c$
cannot vanish. If both $r,s=0$, then $f \in W^2$ and so $W^3 = W^2$.
If both $s,a=0$, then $h$ is proportional to
$q^2$ and $f$ must be of the form
\[\textstyle{\frac{1}{3}}rcq^3 + \alpha q^2 + \beta q\,\!p + \gamma
p^2 + {\rm l.d.t.}\]

\noi But then
$\{f,h\} \propto 2\beta q^2 + 4\gamma q\,\! p
\in W^2$, which forces $\gamma =0.$  Thus $W^3
\subset C^3.$ The canonical transformation
$q \mapsto p,$ $p \mapsto -q$ reduces the subcase with $r,c = 0$ to
the previous one. 

If $W^3 = W^2$ then $W = W^2 \subset P^2$, which contradicts the
assumed maximality of $W$. 

If $W^3 \subset C^3$, then a similar argument shows that $W^4 \subset
C^4$, and so on. Thus $W \subset C$, which again
contradicts the maximality of $W$.
\vskip 6pt

(\emph{ii}) $\dim W_2 = 2$: \ \ Now we may suppose that
$h,g$ form a basis for $W_2$, where $h$ is as above and
\[g(q,p) = rp^2 + spq + tq^2.\]

\noi If $s=0$ then, as $h,g$ are linearly independent, both $p^2,q^2
\in W_2.$ But then $\{p^2,q^2\} = 4pq
\in W$, so that $\dim W_2 =3.$ Without loss of generality, we may
thus assume that $s=1.$

Now $\{h,g\} \in W_2$, and a computation shows that $h,g,\{h,g\}$ are
linearly dependent iff 
\begin{equation}
ac + (at - cr)^2 = 0.
\label{eq:det}
\end{equation}

Again we consider various subcases. If $a=0$ then (\ref{eq:det})
gives $r=0$, and it follows from the above expressions for $h,g$ that
$W_2 = C_2.$ As in case (\emph{i}), the subcase $c=0$ can be reduced
to that of $a=0$ by means of a linear canonical
transformation. It remains to consider the subcase $ac
\neq 0$. We may suppose that $a=1$; (\ref{eq:det}) then implies
that $c < 0.$ Setting $\beta = t-rc$, we may thus take
\[h = p^2 - \beta^2q^2 \;\;\; {\rm  and }\;\;\; \{g,h\} = 2(p^2 +
2\beta pq + \beta^2q^2).\]

\noi as a basis for $W_2$. But now the canonical transformation
\[p \mapsto \frac{1}{\sqrt 2 \beta}(p - \beta q),\;\;\; q
\mapsto
\frac{1}{\sqrt 2}(p + \beta q)\]

\noi reduces this subcase to that of $a=0.$ Thus up to isomorphism we
have $W_2 = C_2$.

Similarly we have $W_k \subseteq C_k$, and so $W \subseteq C.$ The
maximality of $W$ now implies that $W=C.$
\endproof
 
\ms

In particular, the \sa s
$\left\{f(\mu p +q)(p-\mu q ) + g(\mu p + q)\right\},$ where $f,g$
are polynomials and $\mu \in \r$, are all maximal \lsa s of $P(2)$
containing
$P^1(2)$ isomorphic to $C(2)$. (These are the normalizers in
$P(2)$ of the polarizations $\left\{g(\mu p + q)\right\}.$) So
is the ``momentum algebra'' consisting of polynomials which are
at most affine in the position.

\ms

As both $P^2(2)$ and $C(2)$ are quantizable, it follows from
Theorem~\ref{thm:weak} and Corollary~\ref{cor:c} that 
\begin{cor}
Up to isomorphism, $P^2(2)$ and $C(2)$ are the largest
quantizable \sa s of $P(2)$ containing $P^1(2)$.
\label{cor:class}
\end{cor}

Unfortunately, neither Theorem~\ref{thm:max} nor
Corollary~\ref{cor:class} hold in higher dimensions. To see this,
take $n=2$ and consider the \la\
\[\big\{f(q^1)p_1 + g(q^1,q^2,p_2)\big\},\]
 
\noi where $f,g$ are polynomials. This \sa\ is maximal, but not
isomorphic to either $C(4)$ or $P^2(4)$. It is also not
quantizable---if it were, we would obtain a quantization of the
polynomial algebra in $q^2,p_2$, contrary to the Strong No-Go
Theorem. Furthermore, the \sa\ thereof for which $g$ is at most
quadratic in $q^2,p_2$ is maximal quantizable, but also
not isomorphic to either $C(4)$ or $P^2(4)$.

\end{section}


\begin{section}{Discussion}

We have thus completely solved the \gr -\vh\ problem for $\r^2$ in
that we have identified (the isomorphism classes of) the largest
quantizable \lsa s of
$P(2)$ (viz. $P^2(2)$ and $C(2)$) and explicitly
constructed all their possible quantizations (given by (\ref{eq:p2})
and (\ref{eq:sigma}), respectively). It remains to carry out this
program in higher dimensions; the key missing ingredient is a
classification of the maximal \lsa s of $P(2n)$ containing $P^1(2n)$.
Unfortunately, this appears to be a
difficult problem. We emphasize, however, that all the results of
this paper other than Theorem~\ref{thm:max} and
Corollary~\ref{cor:class} hold for arbitrary $n.$

Of course, \gr's classical result is valid only for $\r^{2n}.$
Similar obstructions appear when trying to quantize certain other
phase spaces, e.g.,
$S^2$ and $\T\! S^1$. Complete solutions of
the corresponding \gr -\vh\ problems in these two examples are given
in
\cite{GGH} and \cite{GG1}, respectively. On the other hand, in some
instances there are \emph{no} obstructions to quantization, such as
$T^2$ \cite{Go} and $T^*\!\,\bf R_+$ \cite{GGra}. (Although probably
not of physical interest, it is amusing to wonder what
happens for
$\r^{2n}$, $n>1,$ with an exotic symplectic structure.) It is
important, therefore, is to understand the mechanisms which are
responsible for these divergent outcomes. Already some results have
been established along these lines, to the effect that under certain
circumstances there are obstructions to quantizing both compact and
noncompact phase spaces \cite{GGG,GGra,GG2,GM}. We refer the reader to
\cite{go98} for an up-to-date summary.

\end{section}



\end{document}